\documentclass[journal]{IEEEtran}
\usepackage{booktabs}
\usepackage{amsfonts}
\usepackage{amsmath}
\usepackage[style=ieee,maxnames=4,minnames=3,maxbibnames=3]{biblatex}

\usepackage{array}
\usepackage{mdwmath}
\usepackage{mdwtab}
\usepackage{graphicx,epsfig}
\usepackage{eqparbox}
\usepackage{subfigure}
\usepackage{caption}
\usepackage{mathrsfs}
\usepackage{url}
\usepackage{amssymb}
\usepackage{float}
\usepackage{indentfirst}
\usepackage{tabularx,ragged2e}
\usepackage{cases}
\usepackage{algorithm}
\usepackage{algorithmic}
\usepackage{mathtools}
\usepackage{bm}
\usepackage{setspace}
\usepackage{makecell}
\usepackage{xcolor}
\usepackage{ntheorem}
\usepackage{epstopdf}
\usepackage{verbatim}
\usepackage{enumerate}
\usepackage{color}
\usepackage{subfigure}
\theoremheaderfont{\bfseries}
\theorembodyfont{\upshape}
\theoremseparator{:}

\usepackage{multirow}

\newcolumntype{C}{>{\Centering\arraybackslash}X}

\begin{document}
%\title{\Huge Synergies Between Native Generative AI at the Edge and Scalable Big Model at the Cloud}
\title{\Huge An Edge-Cloud Collaboration Framework for Generative AI Service Provision with Synergetic Big Cloud Model and Small Edge Models}
%\title{\Huge A Generative-AI Service Provision Framework with Synergetic Small Edge Models and Big Cloud Model} 
\author{\IEEEauthorblockN{Yuqing Tian,~\IEEEmembership{Student~Member,~IEEE, }
Zhaoyang Zhang,~\IEEEmembership{Senior~Member,~IEEE, } \\
Yuzhi Yang,~\IEEEmembership{Student~Member,~IEEE, }
Zirui Chen,~\IEEEmembership{Student~Member,~IEEE, }
Zhaohui Yang, ~\IEEEmembership{Member,~IEEE, }
Richeng Jin, ~\IEEEmembership{Member,~IEEE, }
Tony Q.~S.~Quek,~\IEEEmembership{Fellow,~IEEE, }
and Kai-Kit Wong,~\IEEEmembership{Fellow,~IEEE }}
\thanks{This work was supported in part by National Natural Science Foundation of China under Grant U20A20158, National Key R\&D Program of China under Grant 2020YFB1807101, and Provincial Key R\&D Program of Zhejiang under Grant 2023C01021. (\textit{Corresponding Author: Zhaoyang Zhang})}
\thanks{Y.~Tian (e-mail: tianyq@zju.edu.cn), Z.~Zhang (e-mail: ning\_ming@zju.edu.cn), Y.~Yang (e-mail: yuzhi\_yang@zju.edu.cn), Z.~Chen (e-mail: ziruichen@zju.edu.cn), Z.~Yang (e-mail: yang\_zhaohui@zju.edu.cn) and R.~Jin (e-mail: richengjin@zju.edu.cn) are with College of Information Science and Electronic Engineering, Zhejiang University, Hangzhou 310027, China, and also with Zhejiang Provincial Key Laboratory of Info. Proc., Commun. \& Netw. (IPCAN), Hangzhou 310027, China.}
\thanks{T. Q. S. Quek (email: tonyquek@sutd.edu.sg) is with the ISTD Pillar, Singapore University of Technology and Design (SUTD), Singapore 487372, and also with the SUTD-ZJU IDEA Center of Network Intelligence, Singapore 487372.}
\thanks{K.~Wong  (e-mail: kai-kit.wong@ucl.ac.uk) is with the Department of Electronic and Electrical Engineering, University College London, UK.}
}
% \IEEEauthorblockA{College of Information Science and Electronic Engineering, Zhejiang University, Hangzhou, China\\
% Zhejiang Provincial Key Laboratory of Info. Proc., Commun. \& Netw. (IPCAN), Hangzhou, China\\
% }
% E-mail: \{tianyq, ning\_ming, yuzhiyang, yang\_zhaohui, richengjin\}@zju.edu.cn
% }

% make the title area
\maketitle 
\begin{abstract}
%% Generative Artificial Intelligence (GenAI) provides various services to users through content creation, typically involving big AI models (BAIMs). However, the heavy computation and communication overhead poses a challenge to centralized service approaches, as the burden of computing infrastructure is increased, and the reliability and timeliness of long-distance transmission in cloud services cannot be guaranteed. Therefore, decentralizing services from the cloud to the edge is necessary, which establishes native GenAI services to provide private, timely, and personalized experiences. Moreover, edge-cloud collaboration can facilitate collaborative intelligence, which alleviates cloud burdens, improves model quality, and enhances scheme adaptability. （增强方案自适应能力？）Nevertheless, current distributed collaboration solutions face limitations in training BAIMs and deploying native GenAI services, thereby hindering the widespread adoption in the era of large models. To tackle the issues, we propose a bottom-up BAIM architecture, introducing a distributed training framework and a task-oriented deployment solution for BAIMs. The effectiveness of the framework is demonstrated through a use case of image generation. Finally, we outline fundamental research directions to fully exploit the collaborative potential of native GenAI and BAIMs.
    Generative artificial intelligence (GenAI) offers various services to users through content creation, which is believed to be one of the most important components in future networks.
    However, training and deploying big artificial intelligence models (BAIMs) introduces substantial computational and communication overhead.
    %typically involving big AI models (BAIMs), resulting in substantial computational and transmission overhead. 
    This poses a critical challenge to centralized approaches, due to the need of high-performance computing infrastructure and the reliability, secrecy and timeliness issues in long-distance access of cloud services. 
    Therefore, there is an urging need to decentralize the services, partly moving them from the cloud to the edge and establishing native GenAI services to enable private, timely, and personalized experiences. 
    %Edge-cloud collaboration can facilitate collaborative intelligence, alleviate cloud burdens, improve model quality, and enhance adaptability. 
    In this paper, we propose a brand-new bottom-up BAIM architecture with synergetic big cloud model and small edge models, and design a distributed training framework and a task-oriented deployment scheme for efficient provision of native GenAI services. The proposed framework can facilitate collaborative intelligence, enhance adaptability, gather edge knowledge and alleviate edge-cloud burden. 
   % Nevertheless, current distributed solutions face limitations in training BAIMs and deploying native GenAI services, thereby hindering their applications in the era of BAIMs. 
    The effectiveness of the proposed framework is demonstrated through an image generation use case. Finally, we outline fundamental research directions to fully exploit the collaborative potential of edge and cloud for native GenAI and BAIM applications.
    \end{abstract}

\IEEEpeerreviewmaketitle

\section{Introduction}
Generative artificial intelligence (GenAI) is an automated methodology that explores data structures and features to generate content resembling human-created material \cite{AIGC}.
GenAI interacts with users to offer personalized services, including the generation of images, text, and videos. The evolution of GenAI, such as large language models (LLMs) like GPT-4, enhances the quality of service (QoS) and the quality of experience (QoE) in various tasks.
% as they excel in tasks including question-answering, reading comprehension, code generation, and others.
However, the emergent abilities in large-scale GenAI, come at the cost of prohibitive computational and communication resource consumption when operated as centralized cloud service. 
% Currently, the big AI models (BAIMs) \cite{BAIM}, equipped to provide GenAI services, mostly employ a centralized training approach, which is unsustainable due to the distributed data source concern. And the static BAIM architecture designs make the GenAI service provision face challenges, with continuous growth of users, increasing service diversity, and growing application complexity.
Meanwhile, the fifth generation (5G)-advanced towards sixth generation (6G) communication network is shifting from connected intelligence to collaborative intelligence \cite{chen2023foundation}, where the big AI model (BAIM) \cite{BAIM} and small edge models collaborate for service provision. In the anticipated system, the cloud server maintains a unified BAIM by integrating small edge models with diverse tasks. After training, the enhanced BAIM can extract small models corresponding to the tasks, facilitating edge deployment and enabling the delivery of high-performance, low-latency native GenAI services.
% 用户增减 涌现了很多新应用，为了满足用户的不同需求, emerging new applications to deliver to users
% 在未来的通信场景中，云端大模型和边缘小模型之间是共生关系。云服务器通过融合边缘上传的小模型来构建多任务大模型，在不接触原始数据的情况下得到来自边缘和用户的知识。在模型部署阶段，大模型根据功能拆分出应对不同任务的轻量模型，根据用户的需求将模型下发到边缘实现native GenAI。
% 这需要对模型架构设计、大模型的分布式训练方式以及部署方案提出了挑战。
% In the future 6G communication network, big cloud model and small edge models should co-evolve to provide services. Cloud servers construct multi-task BAIMs by integrating small models uploaded from the edge nodes with diverse tasks, obtaining knowledge from the edge and users without accessing the raw data. During the model deployment phase, the BAIM is decomposed into lightweight models tailored for different tasks. These compact small models are then distributed to the edge based on user demands to achieve native GenAI. In essence, the BAIM leverages the advantages of scale and multiple perspectives to deliver high-performance results across multiple tasks. Small models, benefiting from the capabilities output by the large model for individual tasks and reduced scale, can be conveniently deployed at the edge, offering users high-quality, low-latency services.
% In this vision, the scalable architecture, distributed training framework, and the deployment of BAIM need to be intricately designed.
\begin{figure*}[t]
    \centering
    \includegraphics[width=0.92\linewidth]{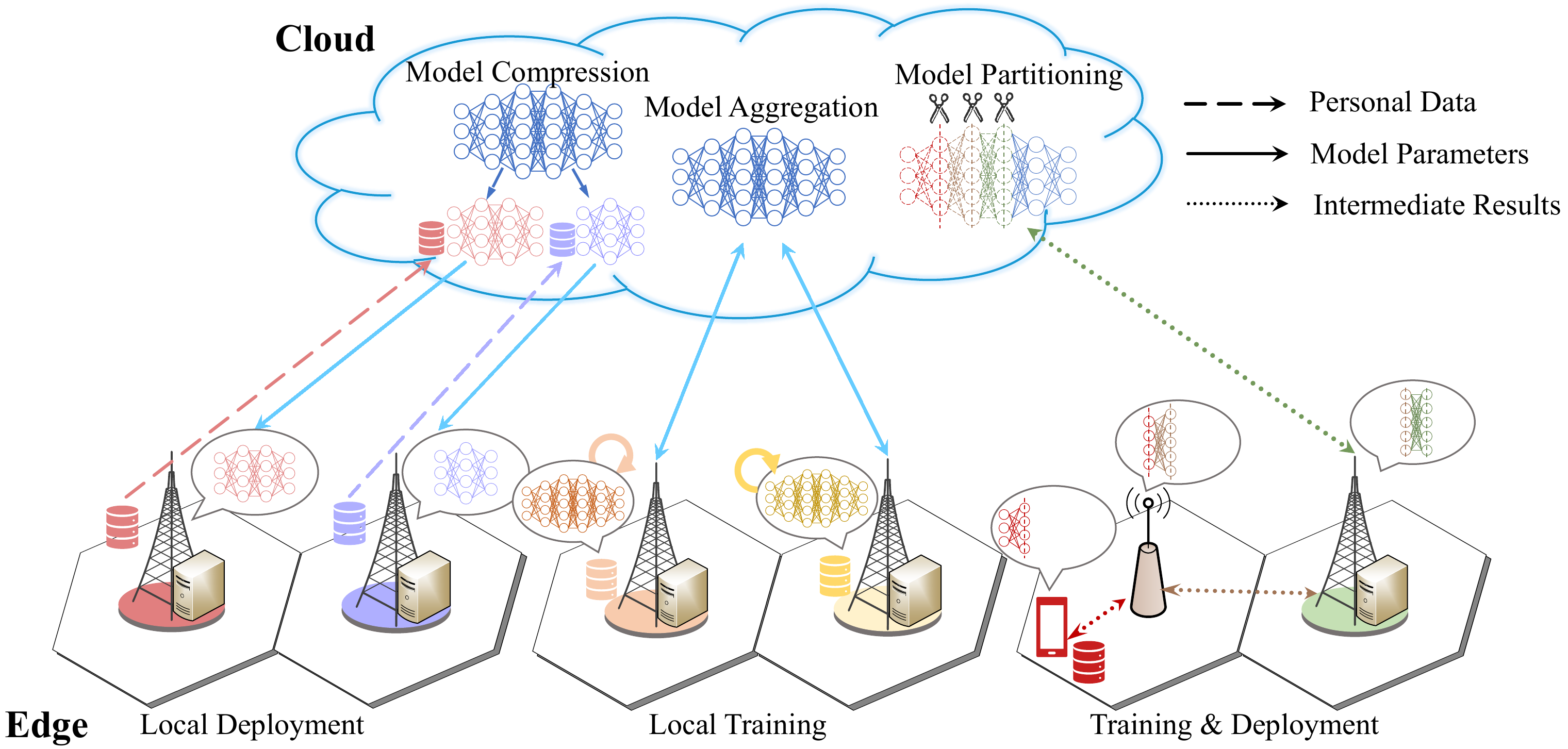}
    \caption{Three frameworks for training and deploying AI models with cloud-edge collaboration. 
    Model compression follows a centralized approach, resulting in smaller models specialized with personal datasets through methods like KD.
    % Model compression employs a centralized approach for training and results in smaller models specialized with personal dataset by KD. 
    Model aggregation merges edge-trained models in an iterative process within the cloud. 
    % Model aggregation involves merging edge-trained models in the cloud in iterative process. 
    % Model partitioning includes jointly training and deploying models by splitting model across different nodes.
    Model partitioning involves the joint training and deployment of models by splitting them across different nodes.}
    \label{compare}
% \vspace{-0.3cm}
\end{figure*}

To tackle issues such as adaptability, edge knowledge acquisition, and cloud overhead, a scalable BAIM architecture with a distributed model training paradigm is required. 

\textit{1) Adaptability:} 
The unified BAIM needs to be capable of addressing the demands of all users. 
To manage the continuous increase in users, service diversity, and application complexity, the BAIM architecture should be scalable.
Additionally, in real-world systems, edges are heterogeneous in communication, computation, and storage capabilities. 
Furthermore, connections between nodes are unstable, so edge nodes may join or leave midway. Therefore, the scalable BAIM should be adaptive for heterogeneous model aggregation and dynamic networks. 

\textit{2) Gathering Edge Knowledge:} 
BAIM has outperformed smaller models in various domains with ample training data.
% and can handle multi-modal input and multi-task scenarios effectively.
In 6G networks, data generated by individual devices is often not enough to train a high-quality model.
To gather them together, edge models could distill local intelligence and transfer it to the center. 
This facilitates the acquisition of global knowledge without direct access to raw data and supports the development of high-quality BAIMs.

\textit{3) Mitigating Cloud Burden:} 
Centralized BAIM training faces increasing demands for data storage, model parameter caching, and computational costs.
The frequent interactions with users amplify the communication burden, posing additional challenges to the central server's capacity.
Simultaneously, the edge network offers substantial computational resources for model training. Distributed BAIM training efficiently leverages edge resources of computing power, storage, and communication, making the process more eco-friendly and cost-efficient.

% Therefore, there is an urgent need to sink GenAI services from the cloud to the edge
% Establishing native GenAI involves deploying the trained BAIM. 

% In this vision, the deployment of large-scale GenAI models is another research focus. 
Moreover, distributed deployment of GenAI services is another research focus, aiming to deliver secure, timely, and personalized services.

\textit{1) Data Security:} Many GenAI services, such as autonomous driving and remote health care, require the collection of real user data. Centralized cloud computing requires users to upload all data to the cloud, raising privacy concerns. 
Deploying native GenAI close to or directly at the data source enables storing data on local servers or user devices, alleviating the need to share sensitive data.

\textit{2) Timeliness of Response:} In contrast to discriminative AI, GenAI generates a tremendous amount of data in response to user requests. Cloud services, relying on long-distance transmission, may suffer from significant latency when delivering these data to users. Native GenAI, with efficient local communication, can enable high throughput and low-latency tasks.

\textit{3) Personalized Services:} 
In response to user requests, edge servers can download a lightweight version of GenAI from the cloud with the necessary functionality, which can be further fine-tuned on the local dataset. Additionally, by grouping users with similar service requirements, the edge can maintain multiple models dedicated to efficiently handling various tasks and applications.

To enhance the QoE and QoS for users in 6G networks, it is crucial to simultaneously leverage the benefits of BAIMs \cite{BAIM} and the advantages of edge services \cite{9311932}. This paper proposes a collaborative scheme that integrates native GenAI with cloud-based BAIM offering a potential solution.
Specifically, we start by analyzing current AI training and deployment strategies in edge-cloud collaboration, demonstrating their limitations. We then summarize the challenges that restrict the distributed training of BAIMs and the deployment of native GenAI. In this context, we propose a bottom-up BAIM architecture, along with a distributed training framework and a task-oriented deployment solution for BAIMs. Within this framework, we demonstrate its contribution in improving service provision through an image generation use case. Finally, we outline the fundamental research directions to fully exploit the potential of native GenAI and BAIM collaboration. 

%% To enhance QoE and QoS for users in 6G networks, it is crucial to leverage the convenience（是指边缘设备能够很方便地被利用？） of edge services and the benefits of large models. This paper aims to tackle this challenge through a collaborative scheme that integrates edge-native GenAI with cloud-based BAIM. First, we analyze current AI training and deployment strategies in edge-cloud collaboration, emphasizing existing limitations. Then we summarize the challenges to achieve the distributed training of BAIMs and the deployment of native GenAI. In this context, we propose a bottom-up BAIM architecture, along with a distributed training framework and a task-oriented deployment solution for BAIMs. Within this framework, we demonstrate its significant contribution to improving user services through a use case of image generation. Finally, we outline the fundamental research directions to fully exploit the potential of native GenAI and BAIM collaboration. 

\begin{table*}[htbp]
  \centering
  \caption{Distributed Model Training and Deployment Frameworks}
  \subtable[Model Configuration and Deployment Phase]{
    \resizebox{\linewidth}{!}{
    \begin{tabular}{|c|c|c|c|c|c|c|c|}
    \hline
    \multirow{2}{*}{Distributed Paradigms} & \multirow{2}{*}{Basic Structure } & \multicolumn{2}{c|}{Model Size} & \multirow{2}{*}{Deployed Model} & \multicolumn{1}{c|}{\multirow{2}{*}{\makecell[c]{Transmission Content \\ in Deployment Phase}}} & \multicolumn{2}{c|}{Inference Latency and Cost} \\
\cline{3-4}\cline{7-8}          &       & Cloud & Edge  &       &       & Comput. & Commun. \\
    \hline
    Knowledge Distillation & Teacher-Student & Big   & Adaptive & Independent and Personalized & - & Low   & - \\
    \hline
    Federated Learning & Top-down & Big   & Big   & Independent &- & High  & - \\
    \hline
    Split Learning & Multi-Partition & Adaptive & Adaptive & Dependent and Cooperative & Intermediate Results & Medium & High \\
    \hline
    Ours  & Bottom-up & Big   & Adaptive & Independent and Personalized & - & Low   & - \\
    \hline
    \end{tabular}%
    }
    }
    
    \subtable[Training Phase]{
         \resizebox{\linewidth}{!}{
        \begin{tabular}{|c|c|c|c|c|c|c|c|}
        \hline
        \multirow{2}{*}{Distributed Paradigms} & \multicolumn{2}{c|}{Data} & \multicolumn{2}{c|}{Training Method} & \multicolumn{1}{c|}{\multirow{2}{*}{\makecell[c]{Transmission Content \\ in Training Phase}}} & \multicolumn{2}{c|}{Training Latency and Cost} \\
\cline{2-5}\cline{7-8}          & Cloud & Edge  & Cloud & Edge  &       & Comput. & Commun. \\
    \hline
    Knowledge Distillation & Personal Data & Personal Data & From Scratch & With Knowledge Logits & Personal Data & High  & High \\
    \hline
    Federated Learning & - & Personal Data & Averaging & From Scratch & Model Parameters & High  & High \\
    \hline
    Split Learning & Labels / - & Personal Data & From Scratch & From Scratch & Intermediate Results & Medium & High \\
    \hline
    Ours  & Common Data & Personal Data & Fine-tuning & From Scratch & Model Parameters & Low   & Low \\
    \hline
          \end{tabular}
    }
    }
  \label{table1}%
\end{table*}%

\section{Overview of model training and deployment with edge-cloud collaboration}
% In this section, we present an overview of the AI model training and deployment frameworks with edge-cloud collaboration in Fig. \ref{compare}. As studied in the 3rd Generation Partnership Project (3GPP) SA1 Release 18 \cite{3GPP}, distributed AI frameworks will play a tremendous role in future wireless networks, including model compression exemplified by knowledge distillation (KD), model aggregation represented by federated learning (FL), and model partitioning also called split learning (SL). We compare these frameworks with our proposed bottom-up BAIM architecture in Table \ref{table1}, highlighting the existing limitations and summarizing the challenges hindering the distributed training and deployment of BAIMs.
In this section, we provide an overview of AI model training and deployment frameworks with edge-cloud collaboration as explored in the 3rd Generation Partnership Project (3GPP) SA1 Release 18 \cite{3GPP}. As illustrated in Fig. \ref{compare}, these distributed AI frameworks include model compression, demonstrated by knowledge distillation (KD), model aggregation, represented by federated learning (FL), and model partitioning, also known as split learning (SL). We compare these frameworks with our proposed bottom-up BAIM architecture in Table \ref{table1}, highlighting existing limitations and summarizing challenges that hinder the distributed training and deployment of BAIMs.

\subsection{Model Compression}
Considering the resource constraints of edge nodes, 
% it is usually impractical to load the entire cloud-trained model onto the edge or user devices for inference. 
loading the entire cloud-trained BAIM onto the edge or user devices for inference is often impractical.
Consequently, model compression, which reduces the model size and computation costs for inference, can partially address the challenge of limited resources. Model compression involves various mature techniques, such as pruning, quantization, low-rank approximation, and KD. 

Taking KD as an example, we describe the lifecycle of model compression technologies from training to deployment.
KD transfers knowledge from a big teacher model to smaller student models based on the same training dataset.
Trained student models are then deployed independently on edge nodes as smaller models are less expensive to evaluate.
% The deployment process involves trained student models being assigned independently on edge nodes, as smaller models are less costly to evaluate.
% KD involves two stages: training a big teacher model and customizing smaller student models. The cloud generates intermediate results with the teacher model, and smaller student models mimic these results, such as the output distribution, hidden states, or attention scores. Trained student models are then deployed independently on edge nodes.
Currently, the application of KD has been extended from classification tasks to generative tasks and has shown remarkable performance on LLMs \cite{KD}. 
% In the context of edge-cloud collaboration, the KD process involves two stages, including training the teacher model and customizing the student model.
% % Initially, the cloud server trains a complex, large teacher model with a substantial number of parameters on a centralized dataset. This teacher model then conducts inference on the training dataset, retaining some middle results. Subsequently, in response to the requests from the edge, the cloud server initializes several smaller, simplified student models based on the edges' states. Each model is trained on the same dataset and the middle results from the teacher, enabling the student to mimic the teacher's output distribution, hidden states, or attention scores. 
% The cloud trains a large teacher model initially, using it to generate intermediate results. In response to edge requests, smaller student models are initialized and trained on the same dataset and teacher's intermediate results. This allows students to mimic the teacher's output distribution, hidden states, or attention scores.
% The trained student models are then deployed independently on the edge nodes.

While model compression significantly improves deployment efficiency in resource-constrained environments, it also introduces several issues, including information loss, poor generalizability, and increased training costs. The training phase introduces a greater burden on the central server and fail to address the problem of distributed data sources.

\subsection{Model Aggregation}
% In distributed learning, 
Model aggregation is another mechanism to integrate information from edge models into a global model. The most representative algorithm is FL. In the FL framework, the cloud server initializes a model and distributes it to each edge node. Then, each edge node conducts model training utilizing its local data. Subsequently, the model parameters are collected and aggregated on the cloud server, formulating a global model. The process is iteratively performed for multiple rounds. Remarkably, this is achieved without sharing raw data among nodes, thereby ensuring data privacy and security. 

FL provides several benefits, including heightened privacy protection, improved edge computing efficiency, and enhanced model generalization capabilities. 
However, the model aggregation is designed for homogeneous models and does not consider the heterogeneity among edge nodes. Furthermore, due to limited resources, the aggregated model is difficult to deploy on devices. Additionally, FL requires frequent exchange of model parameters between the cloud and edge, which becomes impractical for large-scale models. These exchanges are costly and hinder frequent transmissions, particularly under limited communication resources \cite{JSAC21}.
\begin{figure*}[t]
    \centering
    \includegraphics[width=0.86\linewidth]{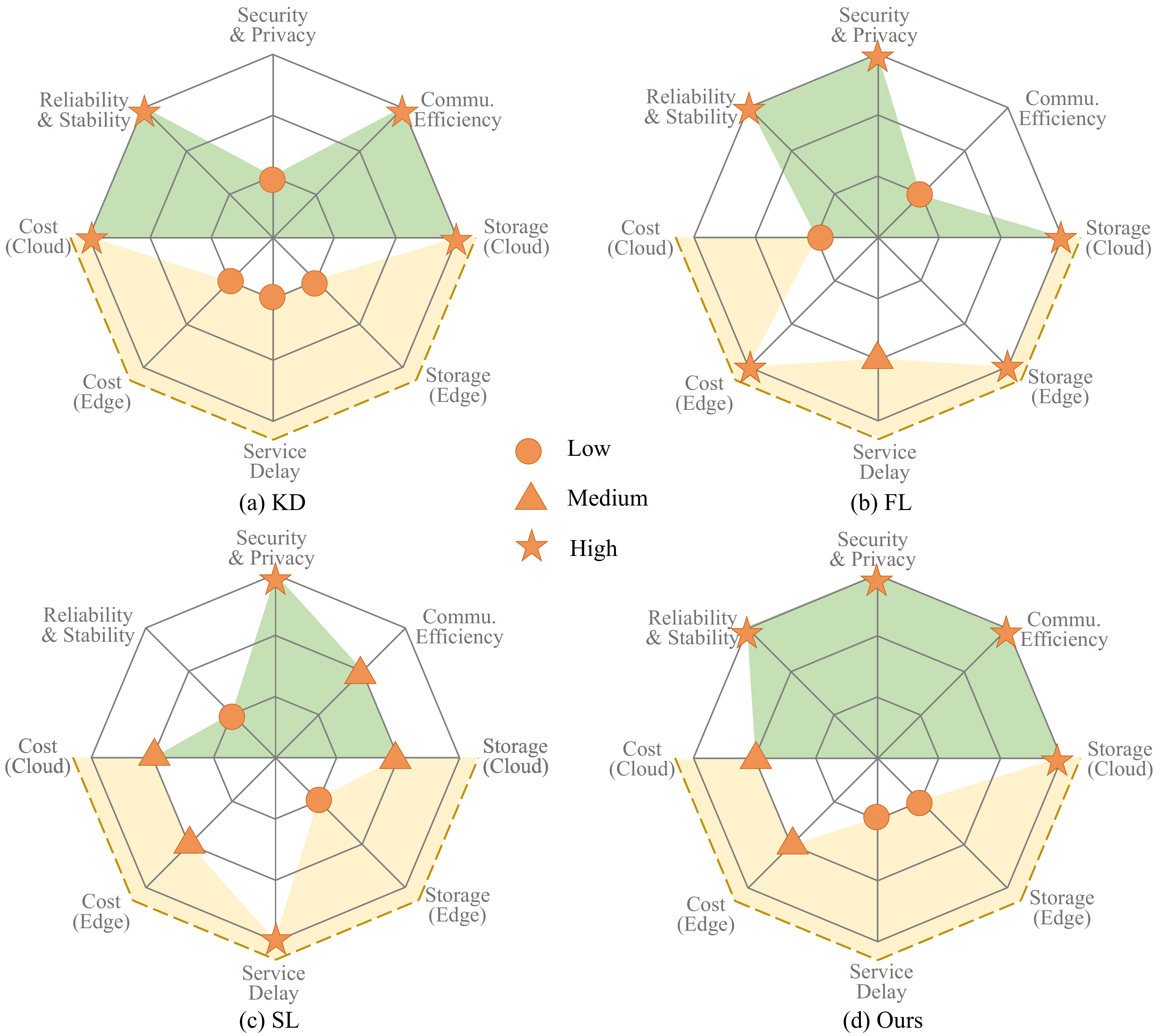}
    \caption{KPI radar chart of four distributed frameworks over edge-cloud networks. 
    % The colored area measures the comprehensive performance of services provided by the framework. 
    }
    % The smaller the area, the better the framework.}
    \label{radar}
% \vspace{-0.3cm}
\end{figure*} 

Federated distillation (FD) provides a solution addressing these challenges. In FD, the users only exchange the models’ intermediate outputs, which are much smaller in size. Each edge node can initialize and train a local model according to its capabilities while storing intermediate results, such as logits in classification tasks. Periodically, these middle results are uploaded to the cloud for aggregation, forming global knowledge. 
% The edge nodes then download this global knowledge , aiming to align the local model outputs with those of the global model. 
The edge nodes then download this global knowledge and incorporate it into their local training, aligning their model outputs with those of the global model.
FD is not constrained by the need to aggregate homogeneous models and effectively reduces communication costs. However, the primary computational load of model training in FD and FL is handled by the edge, placing a significant burden on these nodes. In contrast, the central server's task for aggregation remains relatively simple, leading to an imbalance in computational task distribution.
\begin{figure*}[t]
    \centering
    \includegraphics[width=1\linewidth]{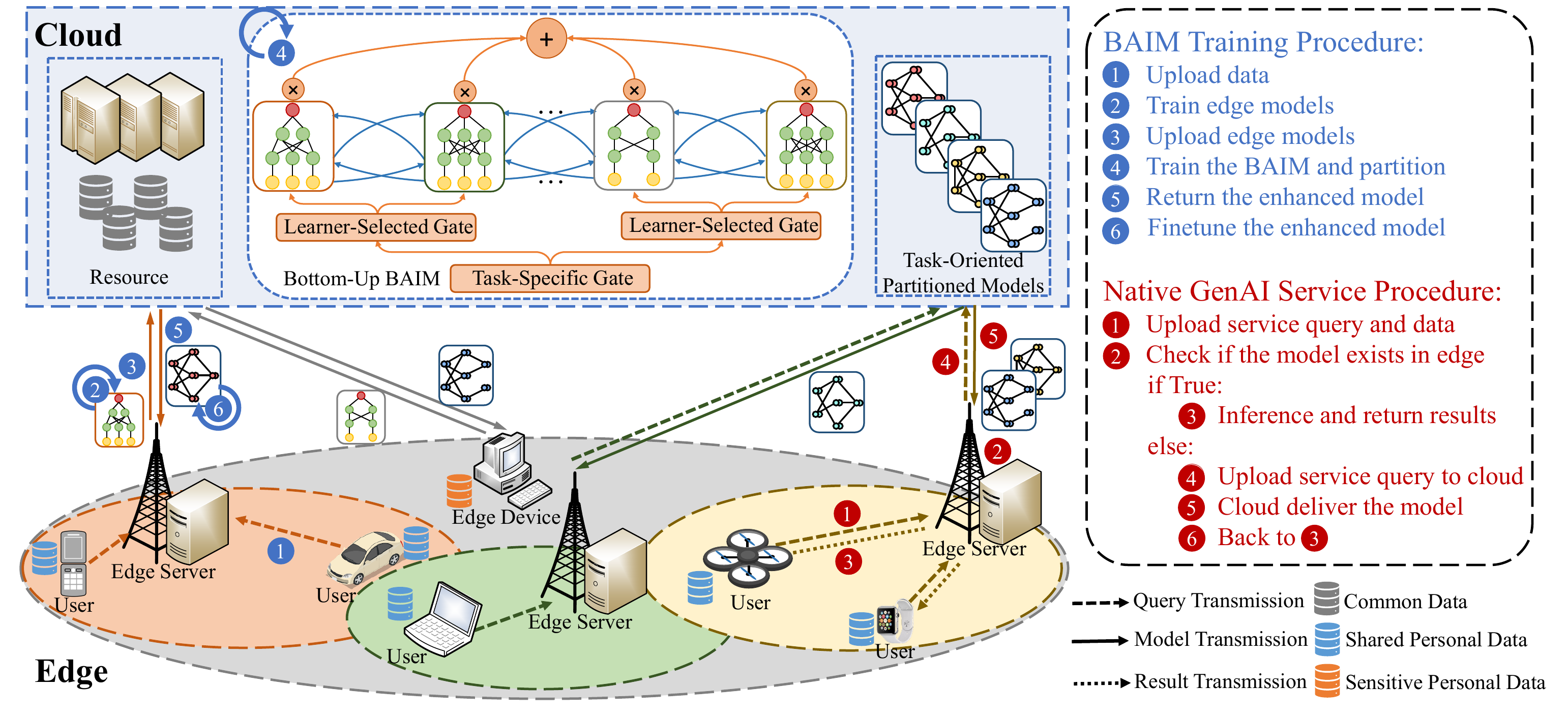}
    \caption{The workflow of our proposed framework with BAIM training and native GenAI service procedures. }
    \label{architecture}
% \vspace{-0.3cm}
\end{figure*}
\subsection{Model Partitioning}
Model partitioning, often referred to as SL, is another method for distributing computational tasks of models. This approach is typically applied in large-scale models that cannot be fully accommodated on a single device or node. In an SL system, the model's structure and parameters are divided into multiple partitions and computed by different nodes within a communication network. This helps balance the computational load across multiple nodes. Moreover, SL does not require users to share their original data, instead, they exchange intermediate results or labels. This approach ensures privacy protection and reduces the communication bandwidth required for exchanging original data \cite{JSTSP}. Generally, SL can be employed both during the model deployment phase and in the model training process.

In the model deployment phase, it is crucial to establish topological partitioning of the well-trained model, considering the following aspects.
%The SL algorithm must consider three pivotal aspects. 
\textit{1) Edge-cloud node information:} This involves considering each node's communication, computational, and storage capabilities. Such information is crucial for determining the optimal distribution and execution of the model across the network.
\textit{2) The size of each layer's output:} 
% an assessment of the dimensions of features produced by each layer post-computation is necessary. This evaluation helps estimate the volume of data required for transmission if a partition occurs at a particular layer given a specific input.
Checking the size of the data generated by each layer is necessary. This helps determine the amount of data that needs to be sent when splitting the model at a specific layer for a given input.
\textit{3) Trade-off between computational cost and communication cost:} 
To reduce communication cost by splitting at layers with smaller outputs, more computation on less powerful devices is often needed. Therefore, achieving a compromise through the loss function is crucial for getting a suitable partitioning solution. 
Previous work employed a joint model split and neural architecture search method to determine the partitioned model \cite{tian2022jmsnas}. This ensures optimal task performance and guaranteed latency within a given communication network.

% \textit{(3) The Key Performance Indicators (KPIs) for QoS and QoE:} 
% KPIs in this context include critical factors such as network inference latency, user energy consumption, and overall service performance. However, these factors may degrade due to inefficiencies and imperfections in data transmission processes.

% Critical factors such as network inference latency, user energy consumption, and the overall performance of services are integral KPIs in this context. However, these may deteriorate due to inefficiencies and imperfections in transmission processes. 

% Integrating SL into the model training phase, coupled with a strategic design of the communication scheme during data transmission, offers a viable solution to mitigate the decline in network performance. 
Engaging in SL during model training and designing communication strategies for data transmission can mitigate the decline in model performance caused by imperfect communication.
By combining the over-the-air computation framework with SL and leveraging the reciprocity of wireless channels, data transmission can be integrated seamlessly into the computation process between model layers. This integration helps reduce the resource expenditure during transmission \cite{JSAC23}. Additionally, combining FL with SL leverages data from various edge nodes. Through cloud-edge collaboration, this approach effectively reduces the computational load of edge nodes, enhancing the overall efficiency of the network \cite{ICLR}. 

\subsection{The Key Performance Indicators (KPIs)}
% The KPIs of the service provision within the edge-cloud collaborative framework include delay, cost, storage, reliability (stability), security (privacy), and communication efficiency.

% Service provision to users within the edge-cloud collaborative framework requires careful consideration of various KPIs. KPIs are used to evaluate the distributed training and deployment framework, and visualized on a radar chart to represent the overall performance of different frameworks, as shown in Fig. \ref{radar}. 
% Service provision to users within the edge-cloud collaborative framework requires careful consideration of various KPIs.
% KPIs are used to evaluate the distributed training and deployment framework, and visualized on a radar chart to represent the overall performance of different frameworks, as shown in Fig. \ref{radar}. 
% Service provision in the edge-cloud collaborative framework involves assessing KPIs, which are evaluated in a radar chart (Fig. \ref{radar}) to depict the comprehensive performance of different frameworks.
Service provision to users within the edge-cloud collaborative framework requires careful consideration of various KPIs.
Fig. \ref{radar} displays six KPIs, including service delay, cost, storage, reliability \& stability, security \& privacy, and communication efficiency. 
% The chart displays six KPIs, favoring lower values for delay, cost, and storage, and higher values for reliability (stability), security (privacy), and communication efficiency. 
Additionally, in a distributed architecture, cost and storage are considered separately for edge and cloud. There is a trade-off between edge and cloud, 
% implying that a decrease in cloud burden often leads to an increase in edge burden. 
% Since cloud resources are often relatively abundant, the cost and storage on the cloud can be regarded as both an overhead and a reflection of task allocation.
% In a distributed framework, a trade-off between edge and cloud involves separate considerations for cost and storage. The cost and storage on the cloud are perceived as both overhead and offloading.
% There is a trade-off between edge and cloud, 
% % implying that a decrease in cloud burden often leads to an increase in edge burden. 
% Since cloud resources are often relatively abundant, the cost and storage on the cloud can be regarded as both an overhead and a computation offloading. 
so the cost and storage on the cloud are placed on the midline of the radar chart, contributing to the evaluation of both system characteristics (green area) and overhead savings (yellow area). 
% The green area represents the system's character, while the yellow area indicates the system's savings. 
% Hence, a smaller orange-colored area corresponds to better system performance.

%% The KPIs of the service provision to users within the edge-cloud collaborative framework include delay, cost, storage, reliability \& stability, security \& privacy, and communication efficiency. 
% Cost and storage 包括在cloud的和在edge的，两者之间存在trade-off。由于云端的计算和存储资源相对丰富，因此可以允许在有一定开销的前提下提升利用率（这两者图中无法体现）
% 上半部分的area表示system的能力（？），下半部分的area表示某个 efficiency

As shown in Fig. \ref{radar}(a), since KD requires users to upload data to the center for model training, it damages the security of the system; Fig. \ref{radar}(b) shows that FL brings huge overhead to the edge due to edge model training, as well as communication inefficiency caused by multiple model transmissions. Fig. \ref{radar}(c) shows that SL causes high service delays due to the transmission of intermediate results and unreliability caused by relying on node connectivity.  Fig. \ref{radar}(d) is the solution we proposed, which has good performance in various KPIs.

\section{A Bottom-Up BAIM Architecture: Distributed Training and Task-Oriented Deployment}
In this section, we introduce the bottom-up BAIM architecture, which leverages edge-cloud collaboration for distributed training and task-oriented deployment.
% leveraging edge-cloud collaboration frameworks to implement its distributed training paradigm and task-oriented deployment strategy.
We first outline the workflow of the framework, encompassing the training process of the BAIM and the lifecycle process of native GenAI services. Subsequently, we describe the architecture, emphasizing its intricate design that enables distributed training and a naturally partitioned deployment scheme. Then we explore its training process in the cloud, which is crucial for generalization with few training data. Finally, we present a deployment strategy based on the task-specific partitioning that empowers native GenAI to dynamically deploy the BAIM on edge nodes. This allows users to obtain performance enhancements of the BAIM and the improved QoE provided by edge services.
\begin{figure*}[t] 
    \centering
    \includegraphics[width=1\linewidth]{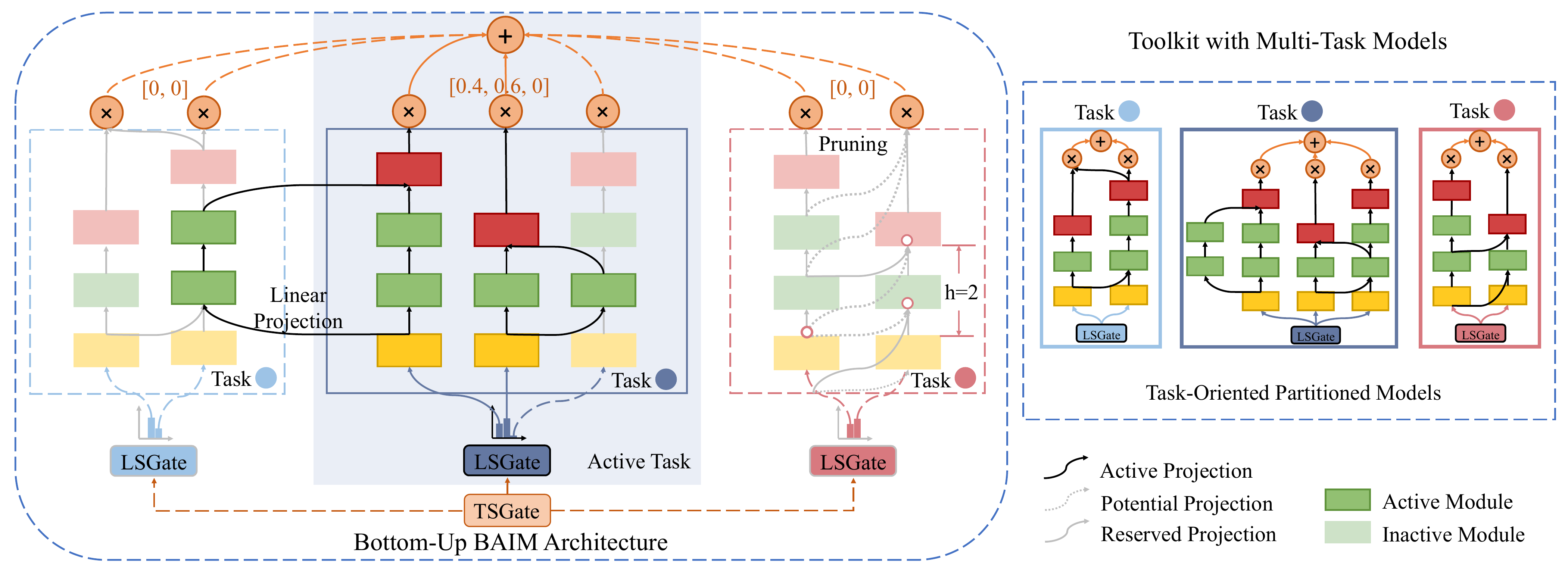}
    \caption{The bottom-up BAIM architecture and task-oriented partitioned models in the toolkit, involving three tasks. The second task is currently selected by TSGate. Dark modules are executed, including top-$k$ ($k=2$) learners chosen by LSGate and modules with linear projection connections to these learners while light modules are inactive in the current round. 
    % Notably, extensive linear connections are observed among learners within and beyond task boundaries. 
    In the third task, gray dashed lines denote the initial potential linear projection (connection height $h=2$) originating from the first learner. During training, pruning filters and reserves a proportion of them, depicted as gray solid lines.}
    \label{model}
% \vspace{-0.3cm}
\end{figure*}
\subsection{Workflow of the Framework}
We depict the workflow of the proposed framework, as illustrated in Fig. \ref{architecture}.
\subsubsection{BAIM training process}
Firstly, users upload shared personal data to the edge, constructing local datasets. Users with privacy-sensitive data act as edge devices, maintaining sensitive personal datasets. Secondly, edge nodes, considering their capabilities and user scale, initialize generative models for respective tasks and train them based on local datasets. Due to distinctive edge characteristics, the trained models may exhibit heterogeneous architectures and features. Next, edge nodes upload the trained models, enabling the cloud to obtain multiple edge models for multi-task and multi-modal learning. The cloud orchestrates edge models with gating neural networks and establishes linear projection connections between stages of different edge models, thereby constructing the bottom-up BAIM architecture. Then, the entire BAIM undergoes training based on the cloud common dataset, achieving superior performance across multiple tasks. Subsequently, the BAIM is easily partitioned based on tasks, yielding compact task-specific models to the edge. Finally, edge nodes can perform personalized fine-tuning on the returned lightweight models using their local datasets.
\subsubsection{Native GenAI service lifecycle}
Firstly, users submit queries to the edge based on their needs, uploading the required service data. Then, the edge checks its local toolbox for requested models. If found, it directly performs inference on user data and returns the results. Otherwise, it requests and downloads the corresponding model from the cloud. If the user service involves sensitive personal data, users can directly acquire the corresponding task model from the cloud.

\subsection{Architecture of the Bottom-Up BAIM}
In the communication systems, the centralized architecture of BAIM poses limitations on acquiring high-quality user data. Inspired by Pathways \cite{PATHWAY} and mixture of experts (MoE) \cite{mustafa2022multimodal}, we propose a bottom-up BAIM Architecture. This architecture maximizes the utilization of user data and expert knowledge extracted by edge models.
Pathways, introduced by Google Mind, represents a next-generation network architecture characterized by multi-tasking, multi-modality, and sparse activation. It is believed that a unified model should be able to expedite learning using existing skills by activating corresponding modules.
MoE, the fundamental structure of GPT-4, combines multiple experts using gating neural networks, enabling adaptive expert output combination. This design harnesses knowledge from diverse experts while mitigating computational demands through sparse gating.
We employ edge models as MoE experts, modularizing the models by establishing linear connections between them. This forms a multi-task, multi-modal, and sparsely activated hierarchical BAIM architecture, as illustrated in Fig. \ref{model}. 

\subsubsection{Multi-task and Gating network}
% The hierarchical gating network (HierGate) not only efficiently manages multitask and multimodal inputs but also boosts computational efficiency through sparsity.
The hierarchical gating network (HierGate) comprises $M$ learner-selected gates (LSGates) and a task-specific gate (TSGate), allowing it to organize multiple tasks and handle multimodal inputs. 
% The introduction of sparsity enhances computational efficiency.
The cloud categorizes $N$ heterogeneous models from edge nodes into task-specific groups, forming $M$ learner squads. Within each group, experts are arranged in parallel, combined through an LSGate. The TSGate is employed to govern the execution of various tasks by routing inputs to the corresponding task. LSGate selects the top $K$ learners most suited to the input, assigning them individual weights. The value of $K$ determines the number of activated learners for a specific task, thereby influencing the computational cost. 
% Typically, K is set to 1 or 2. 
After selecting a task, learners from other tasks do not need to activate the entire model. 
% HierGate achieves efficient structural sparsity, producing an output vector segmented into $M$ sections, with only $K$ non-zero dimensions in an N-dimensional vector. 
HierGate achieves efficient structural sparsity, producing an $N$-dimensional vector segmented into $M$ sections, with only $K$ dimensions being non-zero.
This represents the proportion of outputs from $N$ learners for a specific task, effectively utilizing diverse knowledge from different learners within the same task.
\subsubsection{Modularization and Linear Projection}
Different from typical MoE models that connect learners solely through gating networks, 
% we propose modularizing learners and establishing linear connections between them to facilitate knowledge sharing. 
we suggest organizing learners into modules and establishing linear projection connections among them to facilitate knowledge sharing.
Since learners in the same squad are likely to benefit from each other's expertise, and there could be information associations among learners from different tasks, we create linear connections among $N$ learners following specific rules.
% In Fig. \ref{model}, our rule involves taking the features generated by a learner at stage $i$, applying linear projection to transform them into the feature space of other learners of stage $j$, where $0 \leq j-i \leq d$, and then adding the result to the original input. 
Our rule can be explained as follows: Take the features produced by a learner at stage $i$, perform a linear projection to convert them to the input dimension of other learners at stage $j$ (where $0 \leq j-i \leq h$), and add the result to the original input of stage $j$.
% The hyper-parameter $h$ controls the initial connection density, reflecting the assumption that layers close in proximity process information similarly. 
The hyper-parameter $h \geq 0$ controls the initial connection density, based on the assumption that layers close in depth process original input to a similar extent.
Importantly, connections are established from shallow to deep layers, avoiding the formation of cyclic model structures. Moreover, the dependency among learners can vary significantly. Certain tasks exhibit clear and strong relationships, benefiting from shared features, while others show weaker relations, with shared features less evident. To address this, during model training, we employ pruning to iteratively filter and preserve essential connections. 
% This process reduces the model's parameter size while establishing stable feature-sharing relationships within the model, enabling adaptive knowledge propagation across tasks.
This method reduces the model's parameter size while fostering stable feature-sharing relationships within the model, facilitating adaptive knowledge propagation across tasks.

\subsection{Training BAIM in the Cloud}
For the above model architecture, trainable parameters include gate neural networks, linear connections for feature projection, and each individual learner. Here we introduce three training strategies for BAIM:

\textit{1) Fine-tuning Strategy:} All trainable parameters undergo updates. The uploaded edge models serve as initialization for fine-tuning. This comprehensive adjustment ensures the entire model converges to an optimized configuration, leveraging the knowledge embedded in the locally trained edge models.

\textit{2) Freezing Strategy:} Keep the uploaded edge models unchanged, only update the connections between models and the gate network. This maintains the individuality of each learner throughout the training process, serving as static contributors to the overall model.

\textit{3) Scratch Strategy:} Train all parameters starting from random initialization. This approach allows for a thorough evolution of the entire model architecture, emphasizing the independence of the unified BAIM from pre-existing knowledge encapsulated in the uploaded edge models.
\begin{figure*}[t] 
    \centering
    \includegraphics[width=0.98\linewidth]{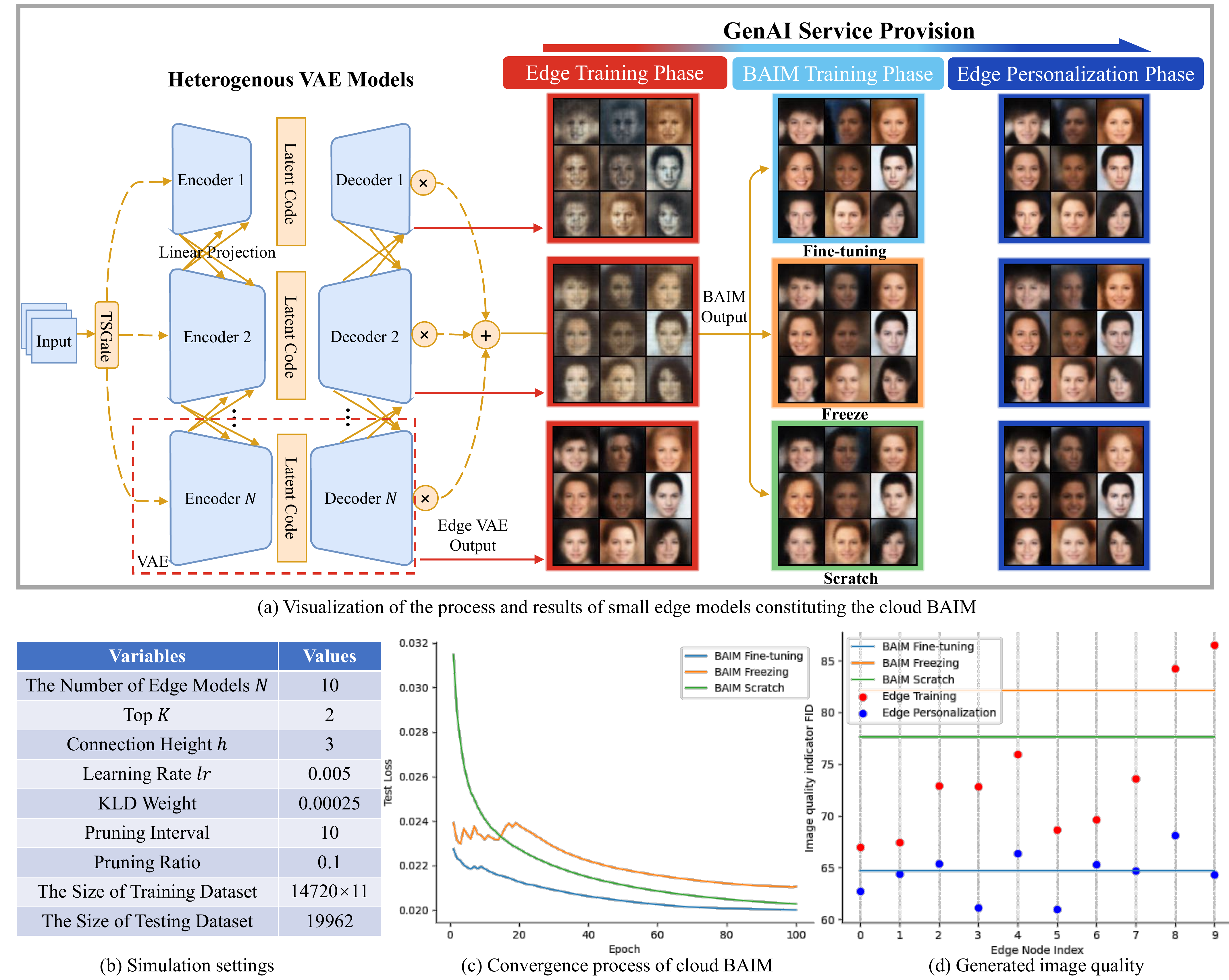}
    \caption{The case study on image generation service provision.}
    \label{simulation}
% \vspace{-0.3cm}
\end{figure*}

In addition, during the model training process, it is crucial to consider knowledge online updating solutions from dynamic environments, particularly from edge nodes. This involves the following three approaches:

\textit{1) Continual Learning:}
Continual learning refers to the model acquiring new capabilities without forgetting original tasks. Unlike multitask learning \cite{chen2023mod}, where all tasks are learned simultaneously, continual learning involves a gradual increase in tasks. In edge-cloud collaborative networks, as the number of users served by edge nodes continues to grow, and their required tasks become more diverse, edge nodes continuously upload models to participate in the aggregation of the unified model. Our bottom-up BAIM is a scalable architecture that, for scenarios with additional learners, can fine-tune the gate network to learn new tasks gradually. 
% However, when the accumulation of new learners reaches a certain threshold, a global fine-tuning of the model is necessary to prevent it from becoming outdated.

\textit{2) Model-Level Pruning:}
Model-level pruning involves trimming sub-models from a large model. In the scalable BAIM, as the number of edge-connected models increases, the quantity of learners for each task grows continuously. Simultaneously, some poorly performing learners are rarely or almost never activated by LSGate. For these learners, in model-level pruning, the parts that do not assist other learners can be removed, and the modules that assist other learners are directly merged into the corresponding learners. This ensures the storage efficiency and computational effectiveness of BAIM.

\textit{3) Few-Shot Learning:}
Few-shot learning refers to training models to generate qualified context or make accurate predictions when provided with very limited examples \cite{lin2023multimodality}. For cloud-based BAIM, the data accessible directly is typically generic public data, and for different tasks, there may be only a few shot samples or even zero shots. BAIM needs to adjust model parameters with limited samples to ensure good performance across various tasks. This requires thoroughly exploring the correlations between different tasks and leveraging the knowledge shared among learners.

These three knowledge-updating strategies are crucial for the self-maintenance of a long-term deployment framework that needs continuous evolution and adaptation to changing conditions. The bottom-up BAIM architecture inherently provides corresponding solutions to these challenges.

\subsection{Native GenAI with Task-Oriented Deployment}
In addition to the unified multi-task BAIM obtained on the central server, the framework can also implement model compression and model partitioning to obtain enhanced performance for different tasks. These compact and lightweight models can be deployed to edge nodes, providing users with native GenAI services.
As mentioned in Section II, model compression typically results in a loss of model performance or requires additional fine-tuning. However, the architecture we propose has unique properties that allow the decomposition of the model into compact models for corresponding tasks based on TSGate without sacrificing performance. Connections involving shared experiences are replicated into new models, as shown in Fig. \ref{model}. The resulting structure for each task is a MoE model architecture with linear connections, each having its own LSGate for selecting learners conditioned on the input, demonstrating excellent generalization capabilities for the task.
As a result, we obtain $M$ compact models. For edge nodes, these lightweight models can be downloaded from the cloud as needed, meeting user service requirements. This approach enables edge services to achieve performance comparable to the cloud while harnessing the advantages of edge services.

\section{A Case Study: Image Generation Service Provision}
In this section, we demonstrate a typical image generation service with variational autoencoder (VAE) models and showcase image quality improvement through edge-cloud collaboration in our framework. 
The simulation setup, as illustrated in Fig. \ref{simulation} (b), includes 10 edge nodes training their heterogeneous VAE models using diverse local datasets and subsequently evaluating them on a common test dataset derived from the CelebA dataset. After local training, edge models are uploaded to the cloud, where BAIM is trained using three different strategies: finetuning, freezing, and scratching. Finally, the model obtained by the finetuning strategy is sent back to the edge for further personalized fine-tuning. The testing samples of three phases are illustrated in the three columns of images in Fig. \ref{simulation} (a), respectively.

Fig.\ref{simulation} (c) shows the convergence of the testing loss under the three training strategies for BAIM. The finetuning strategy exhibits the best convergence and performance, while the freeze strategy initially oscillates for the first several epochs before converging, with the worst loss function performance. Due to random initialization, the scratch strategy starts with a high initial loss, experiences a rapid decrease, and reaches the middle loss level. This is attributed to the lack of knowledge extracted by the edge model in the scratch approach, preventing it from achieving the performance of finetuning despite having the same model size and parameter space.

Fréchet Inception Distance (FID) is a widely used metric for evaluating the quality of generated images. 
% It quantifies the similarity between the distributions of real and generated images, with a lower FID indicating a closer match between them. 
It measures the similarity between the distribution of real images and generated images. A lower FID implies that the generated images closely match the real ones.
Fig.\ref{simulation} (d) illustrates images generated by models under three BAIM training strategies (represented by three horizontal lines) and images generated by the edge during initial edge training and edge personalization after BAIM deployment to the edge (represented by two types of scatter). Notably, the fine-tuning strategy proves effective, significantly improving image quality compared to the original edge model. 

\section{Challenges and Potential Research Opportunities}
Our proposed framework paves the way for efficient service provision in 5G-advanced and 6G communication networks. However, it introduces challenges that demand attention. In this section, we analyze challenges in data management, model fusion scheme design, and node management, which provide potential research directions.

\subsection{Data Management}
We address data privacy concerns by distinguishing between sensitive personal data, shared personal data, and common data. Moving forward, a comprehensive solution for data management and generation is yet to be developed. 
% Here are two prospective directions:
\subsubsection{Secure Data Management Scheme}
Vigorous safeguards should be established to protect data during storage and transmission. This includes end-to-end data encryption and enhanced identity verification and authorization mechanisms. Additionally, anonymization and desensitization techniques for cloud-side common data to minimize the risk of information leakage are also expected.
\subsubsection{Substituting User Raw Data with Synthetic Data}
This involves the application of differential privacy techniques, generative adversarial networks (GANs), and data perturbation methods to generate synthetic data with authentic data features. In light of the advancements in artificial intelligence generated content (AIGC), synthetic data could serve as a more secure and reliable alternative for model training data.

\subsection{Model Fusion Scheme}
During the model training phase, designing more promising model fusion strategies and asynchronous update mechanisms could lead to sustained improvements in computing performance and communication efficiency.

\subsubsection{Optimizing Heterogeneous Architecture Fusion Strategies}
Edge models vary in structure concerning depth and width,  as well as in architectures like convolutional neural networks (CNNs), recurrent neural networks (RNNs), and Transformers. For efficient multi-task learning with these diverse models, it's crucial to improve heterogeneous architecture fusion strategies, including projection methods, connecting rules and pruning strategies. This enables better exploration of task correlations and information sharing for mutual reinforcement.

\subsubsection{Designing Asynchronous Update Mechanisms}
Asynchronous update systems allow immediate uploads from each edge node once their calculations are complete, effectively reducing waiting times. As edge models are independently trained, the cloud consistently receives these well-trained small models and updates the BAIM.  A delicately designed asynchronous update mechanism is required to orchestrate the fusion of these edge models. This mechanism should strike an optimal balance between the BAIM's staleness and computational cost.

\subsection{Node Management}

Node management involves the flexible monitoring, adjustment, and coordination of changes within a distributed system. Effective node management can enhance the stability and reliability of the system, thus reducing any performance decline resulting from node anomalies.

\subsubsection{Adapting Dynamic Edge Networks}

Edge networks in practical systems constantly change, leading to possible instability in edge nodes, including their access and disconnection. Systems must adapt to add new nodes and handle failed ones, and respond to changes in current nodes' states. Given edge nodes are often distributed and can be mobile, the system needs great adaptability to tackle issues like network delays, data loss, or node availability changes. Proper node management plays a crucial role in model evolution and requires well-planned enhancement mechanisms.

\subsubsection{Addressing Security Threats}

In open cloud-edge systems, malicious node attacks are an inevitable issue. These can potentially include dishonest nodes sabotaging model performance through false training results, or disrupting system operation via denial-of-service (DoS) attacks. To counter these, security measures involve tamper and anomaly detection, trust assessment, and cross-verification of node updates against historical behavior or peer updates. Implementing a trust-based reputation system to guide node interactions is also fundamental.

% \subsection{Resource allocation and scheduling}

% Resource allocation and scheduling are crucial, especially in dynamic edge networks. Optimizing system performance and resource efficiency requires the effective and fair allocation of resources among edge nodes through scheduling.
% \subsubsection{computing resource}
% Computing resource allocation needs to be tailored to each node's specific characteristics. During model training, nodes process tasks independently, necessitating resource allocation alignment with their workloads. An adaptive strategy dynamically adjusts resource distribution based on edge nodes' demands and workloads. Additionally, exploring mechanisms for estimating and predicting node resource needs can further enhance effective resource management in the system.
% \subsubsection{communication resource}
% Exploring advanced scheduling algorithms for dynamic communication resource allocation is crucial.  For instance, critical task data transmission should be prioritized during network congestion. Machine learning methods can enhance resource scheduling by predicting future communication demands based on historical patterns. Furthermore, addressing the mobility of edge nodes necessitates a mechanism such as Adaptive Network Coding (ANC) that adapts to network changes, ensuring efficient data transmission across diverse environments.

\section{Conclusion}
In conclusion, the synergies between edge-native GenAI and cloud-based BAIMs emerge as a crucial component in 6G communication networks, promising elevated QoE and QoS. The presented framework strategically tackles prevailing constraints in AI training and deployment, with a particular focus on mitigating challenges associated with distributed training of BAIMs and the deployment intricacies of native GenAI. The framework's showcased effectiveness in an image generation use case underscores the substantial potential of this collaborative paradigm. Furthermore, comprehensive research directions are delineated to unlock the full spectrum of possibilities in the synergy between native GenAI and BAIMs.

% \subsection{Extension to Heterogeneous NN frameworks}
% Transformer
% \subsection{Influence of Hyperparameters}
% high model-level pruning ratio, large single network / low pruning, small network
% \subsection{}
\printbibliography

@article{AIGC,
  title={Unleashing the power of edge-cloud generative {AI} in mobile networks: A survey of {AIGC} services},
  author={Xu, M. and Du, H. and Niyato, D. and others},
  journal={arXiv preprint arXiv:2303.16129},
  year={2023}
}

@article{BAIM,
  title={Big {AI} models for {6G} wireless networks: Opportunities, challenges, and research directions},
  author={Chen, Z. and Zhang, Z. and Yang, Z.},
  journal={arXiv preprint arXiv:2308.06250},
  year={2023}
}

@ARTICLE{JSAC21,
  author={Chen, M. and Gündüz, D. and Huang, K. and others},
  journal={IEEE Journal on Selected Areas in Communications}, 
  title={Distributed Learning in Wireless Networks: Recent Progress and Future Challenges}, 
  year={2021},
  volume={39},
  number={12},
  pages={3579-3605},
  doi={10.1109/JSAC.2021.3118346}}

@ARTICLE{JSTSP,
  author={Xu, W. and Yang, Z. and Ng, D. W. K. and others},
  journal={IEEE Journal of Selected Topics in Signal Processing}, 
  title={Edge Learning for {B5G} Networks With Distributed Signal Processing: Semantic Communication, Edge Computing, and Wireless Sensing}, 
  year={2023},
  volume={17},
  number={1},
  pages={9-39},
  doi={10.1109/JSTSP.2023.3239189}}

@article{3GPP,
  title={An Overview of the {3GPP} Study on Artificial Intelligence for {5G} New Radio},
  author={Lin, X.},
  journal={arXiv preprint arXiv:2308.05315},
  year={2023}
}

@article{KD,
  title={Knowledge Distillation of Large Language Models},
  author={Gu, Y. and Dong, L. and Wei, F. and Huang, M.},
  journal={arXiv preprint arXiv:2306.08543},
  year={2023}
}

@ARTICLE{JSAC23,
  author={Yang, Y. and Zhang, Z. and Tian, Y. and others},
  journal={IEEE Journal on Selected Areas in Communications}, 
  title={Over-the-{Air} Split Machine Learning in Wireless {MIMO} Networks}, 
  year={2023},
  volume={41},
  number={4},
  pages={1007-1022},
  doi={10.1109/JSAC.2023.3242701}}

@inproceedings{ICLR,
  title={{MocoSFL}: enabling cross-client collaborative self-supervised learning},
  author={Li, J. and Lyu, Lingjuan and Iso, Daisuke and Chakrabarti, Chaitali and Spranger, Michael},
  booktitle={The Eleventh International Conference on Learning Representations},
  year={2022}
}

@Article{PATHWAY,
author = {Google}, 
title = {{Introducing Pathways: A next-generation {AI} architecture}}, 
note = {https://blog.google/technology/ai/introducing-pathways-next-generation-ai-architecture/.},}

@article{mustafa2022multimodal,
  title={Multimodal contrastive learning with {LiMoE}: the language-image mixture of experts},
  author={Mustafa, B. and Riquelme, C. and Puigcerver, J. and others},
  journal={Advances in Neural Information Processing Systems},
  volume={35},
  pages={9564--9576},
  year={2022}
}

@inproceedings{lin2023multimodality,
  title={Multimodality helps unimodality: Cross-modal few-shot learning with multimodal models},
  author={Lin, Zhiqiu and Yu, Samuel and Kuang, Zhiyi and Pathak, Deepak and Ramanan, Deva},
  booktitle={Proceedings of the IEEE/CVF Conference on Computer Vision and Pattern Recognition},
  pages={19325--19337},
  year={2023}
}

@article{chen2023foundation,
  title={Foundation Model Based Native {AI} Framework in {6G} with Cloud-Edge-End Collaboration},
  author={Chen, Xiang and Guo, Zhiheng and Wang, Xijun and Yang, Howard H and Feng, Chenyuan and Su, Junshen and Zheng, Sihui and Quek, Tony QS},
  journal={arXiv preprint arXiv:2310.17471},
  year={2023}
}

@ARTICLE{9311932,
  author={Xiao, Yong and Shi, Guangming and Li, Yingyu and Saad, Walid and Poor, H. Vincent},
  journal={IEEE Communications Magazine}, 
  title={Toward Self-Learning Edge Intelligence in {6G}}, 
  year={2020},
  volume={58},
  number={12},
  pages={34-40},
  doi={10.1109/MCOM.001.2000388}}

@inproceedings{chen2023mod,
  title={Mod-Squad: Designing Mixtures of Experts As Modular Multi-Task Learners},
  author={Chen, Zitian and Shen, Yikang and Ding, Mingyu and Chen, Zhenfang and Zhao, Hengshuang and Learned-Miller, Erik G and Gan, Chuang},
  booktitle={Proceedings of the IEEE/CVF Conference on Computer Vision and Pattern Recognition},
  pages={11828--11837},
  year={2023}
}

@inproceedings{tian2022jmsnas,
  title={{JMSNAS}: Joint model split and neural architecture search for learning over mobile edge networks},
  author={Tian, Yuqing and Zhang, Zhaoyang and Yang, Zhaohui and Yang, Qianqian},
  booktitle={2022 IEEE International Conference on Communications Workshops},
  pages={103--108},
  year={2022},
  organization={IEEE}
}
\end{document}